\documentclass[aps,prl,amsfonts,twocolumn,groupedaddress]{revtex4}

\usepackage{graphics,epsfig}

\newcommand {\be} {\begin{equation}}
\newcommand {\ba} {\begin{eqnarray}}
\newcommand {\ee} {\end{equation}}
\newcommand {\ea} {\end{eqnarray}}

\begin{document}

\preprint{WM-05-123}

\title{New experimental constraints on polarizability
corrections to hydrogen hyperfine structure}

\author{Vahagn Nazaryan$^{(a,b)}$, Carl E.\ Carlson$^{(b)}$, and Keith A.\ Griffioen$^{(b)}$}
\affiliation{
$^{(a)}$
Department of Physics, Hampton University, Hampton,
VA 23668\\
$^{(b)}$Department of Physics, College of William and Mary, Williamsburg, VA 23187, USA }

\date{\today}

\begin{abstract}

We present a state-of-the-art evaluation of the polarizability
corrections---the inelastic nucleon corrections---to the hydrogen
ground-state
hyperfine splitting using analytic fits to the most recent
data.  We find a value $\Delta_{\rm pol} = 1.3 \pm 0.3$ ppm.
This is 1--2 ppm smaller than the value of $\Delta_{\rm pol}$ deduced
using hyperfine splitting data and elastic nucleon corrections obtained
from modern form factor fits.

\end{abstract}

\maketitle




Hyperfine splitting in the hydrogen ground state is measured
to 13 significant figures~\cite{Karshenboim:1997zu},
    \be
    E_{\rm hfs}(e^-p) = {\rm 1\ 420.405\ 751\ 766\ 7  (9)\ MHz} \,.
    \ee
Theoretical understanding of hydrogen hyperfine splitting is
far less accurate, being at about the part-per-million (ppm)
level. The main theoretical uncertainty lies in proton
structure corrections, which are not presently calculable
from fundamental theory. However, proton structure corrections can be calculated as functionals of quantities measurable in
other experiments, specifically as integrals~\cite{Zemach}
over proton form factors measured in elastic electron-proton
scattering plus
integrals~\cite{Iddings,Drell:1966kk,DeRafael:mc,Gnaedig:qt,Faustov:yp}
over structure functions measured in inelastic polarized electron-proton scattering. The quality of the data
for the latter has improved greatly in recent times,
especially in the lower momentum transfer region which is
important for proton hyperfine corrections.  In this article,
we present a state-of-the-art evaluation of the polarizability
corrections using analytic fits to the most recent data, in
particular using lower momentum transfer
data~\cite{Fatemi:2003yh,models,deur} from Jefferson Lab.

The proton structure corrections, $\Delta_S$ below, can be
isolated by taking the experimental values for the hyperfine
splitting (hfs) and subtracting the other corrections, which include
QED corrections $\Delta_{\rm
QED}$, recoil corrections $\Delta_R^p$, as well as some
smaller terms due to hadronic vacuum polarization $\Delta_{\rm
hvp}^p$, muonic vacuum polarization $\Delta_{\mu{\rm vp}}^p$,
and weak interactions $\Delta_{\rm weak}^p$.  One
has~\cite{Bodwin:1987mj,Volotka:2004zu,dupays}
\ba \label{eq:Ephfs}
    E_{\rm hfs}(e^-p) &=&
    \big (1+\Delta_{\rm QED}+\Delta_R^p+\Delta_{\rm hvp}^p
        \nonumber \\[1ex]
    && +\ \Delta_{\mu{\rm vp}}^p+\Delta_{\rm weak}^p+\Delta_S \big)
    \, E_F^p \,,
    \ea
where the scale is set by the Fermi energy, given by
\be \label{eq:Fermi}
    E_F^p=\frac{8}{3\pi}\alpha^3 \mu_B\mu_p
    \frac{m_e^3 m_p^3}{(m_p+m_e)^3}
    \ee
for an electron of mass $m_e$ bound to a proton of mass $m_p$, magnetic
moment $\mu_p=(g_p/2) (e/2m_p)$, Land\'e $g$-factor $g_p$, and fine
structure constant $\alpha$.
By convention, the exact magnetic moment $\mu_p$ is used for
the proton, but only the lowest order term, the Bohr magneton
$\mu_B$, is inserted for the $e^-$.

The proton structure
correction inferred from atomic hyperfine splitting data by
removing other corrections is
\begin{equation}
\Delta_S = \Delta_Z + \Delta_{\rm pol} = -38.58(16){\rm\ ppm}.
\end{equation}
The uncertainty comes mainly from $\Delta_R^p$; we used $5.84\pm 0.15$ ppm. The central value~\cite{Volotka:2004zu} uses the CODATA~\cite{Mohr:2000ie} charge radius and the dipole magnetic radius.  The uncertainly encompasses a spread due to other choices of these radii.  A discussion is in~\cite{Brodsky:2004ck}. Other quantities were taken from~\cite{Volotka:2004zu}. 
Eliminating the large QED corrections by
using both hydrogen and
muonium hfs~\cite{Brodsky:2004ck} leads to a similar result, with an uncertainty limit 0.18 ppm.

\vglue - 11.3 cm          \hfill hep-ph/0512108v3.x              \vglue 10.89cm

The structure-dependent correction $\Delta_S$ is conventionally split
into two terms, $\Delta_Z$ and $\Delta_{\rm pol}$.
The bulk of the first term was calculated by
Zemach~\cite{Zemach} long ago, and the modern expression is
\begin{equation}
\Delta_Z = - 2 \alpha m_e  r_Z
    \left( 1 + \delta_Z^{\rm rad} \right) \,,
\end{equation}
where the Zemach radius is
\begin{equation}
 r_Z
    = - \frac{4}{\pi}\int_0^\infty\frac{dQ}{Q^2}
    \left[  G_E(Q^2) \frac{G_M(Q^2)}{1+\kappa_p} -1\right] \,;
\label{Zemachint}
\end{equation}
$G_E$ and $G_M$ are the electric and magnetic form factors of
the proton, normalized so that $G_E(0) = G_M(0)/(1+\kappa_p) =
1$,  $\kappa_p=(g_p-2)/2$. The radiative correction $\delta_Z^{\rm rad}$ is estimated
in~\cite{Bodwin:1987mj} and calculated
in~\cite{Karshenboim:1996ew} for form
factors that are represented by dipole forms:
 $\delta_Z^{\rm rad}
= (\alpha/3\pi)
    \left[ 2 \ln ( \Lambda^2/ m_e^2 ) - 4111/420 \right]$.
For $\Lambda^2 = 0.71$ GeV$^2$, one finds $\delta_Z^{\rm rad}
= 0.0153$; for a $\Lambda^2$ corresponding to the relatively large charge radius in~\cite{Sick:2003gm}, one would instead get 0.0150, or a 0.01 ppm change in the hyperfine splitting.

The second term, $\Delta_{\rm pol}$, involves contributions
where the proton is excited~\cite{Iddings, Drell:1966kk, DeRafael:mc,
Gnaedig:qt, Faustov:yp}.  In the limit where the electron
mass is neglected except for one overall factor,
\be
    \Delta_{\rm pol}=\frac{\alpha m_e}{\pi g_p m_p}
    (\Delta_1+\Delta_2),
    \ee
[the prefactor is 0.2265 ppm] with
\ba
  \Delta_1 &=& \frac{9}{4}\int_0^\infty \frac{dQ^2}{Q^2}\left\{F_2^2(Q^2) + \frac{8m_p^2}{Q^2}
B_1(Q^2)\right\}   \,,
										\nonumber\\
\Delta_2 &=& -24m_p^2\int_0^\infty \frac{dQ^2}{Q^4}B_2(Q^2) .
	\label{eq:Delta}
    \ea
Here
$F_2$ is the Pauli form factor
of the proton,
\ba
B_1 = \int_0^{x_{\rm th}} dx \, \beta(\tau) g_1(x,Q^2)  \,,  \nonumber \\
B_2 = \int_0^{x_{\rm th}} dx \, \beta_2(\tau) g_2(x,Q^2)  \,,
\label{eq:B}
\ea
\ba
    \beta(\tau) &=& \frac{4}{9} \left(
        - 3\tau + 2\tau^2 + 2(2-\tau)\sqrt{\tau(\tau+1)}  \right)\quad{\rm and}
                \nonumber \\
\beta_2(\tau) &=& 1+2\tau-2\sqrt{\tau(\tau+1)}  \,,
    \ea
in which $\tau = {\nu^2} \big/ {Q^2}$, $\nu$ is the lab-frame energy transfer, $Q^2$ is the squared 4-momentum
transfer, $x_{th} = Q^2/(2m_p m_\pi + m_\pi^2 + Q^2)$, $m_\pi$ is the pion mass, and $g_1$ and $g_2$ are the spin-dependent structure functions,  measured in doubly polarized electron  proton inelastic scattering.

Ref.~\cite{Faustov:yp} was to our knowledge the first to
use $g_{1,2}$ data to obtain results not consistent
with zero for the polarizability
corrections. However, $\Delta_1$ and
$\Delta_2$  are sensitive to the behavior of the structure
functions at low $Q^2$. Hence with new low $Q^2$
data available~\cite{Fatemi:2003yh,deur}, there is a
significant possibility that the numerical results for
$\Delta_{\rm pol}$ could require noticeable revision.




The integrands of Eq.~(\ref{eq:Delta}) converge
because there are factors that cancel the poles at $Q^2 = 0$. The
function $\beta_2$ has limiting behavior 
\ba 
\beta_2(\tau) &=& \left\{
            \begin{array} {cl}
            1/4\tau = Q^2/4\nu^2  \qquad & \tau \to \infty, \ Q^2 \to 0  \\
            1 \qquad & \tau \to 0, \ Q^2 \to \infty  \,.
            \end{array}
            \right.   
\ea

\noindent 
Given that $\nu$ is never zero for inelastic processes, even for
$Q^2\to 0$, and noting that the width of the integration region for the
$B_i$ is proportional to $Q^2$ for small $Q^2$, one sees that the
integral for $\Delta_2$ is well behaved for finite $g_2$.

For the integral $\Delta_1$ to be finite, given
\ba
\beta(\tau) &=&
\left\{
            \begin{array} {cc}
            1 -5/{18\tau}\qquad & \tau \to \infty  \\
            0 \qquad & \tau \to 0
            \end{array}
            \right.  \,,
\ea
one needs a cancellation that follows from the Gerasimov-Drell-Hearn
(GDH)~\cite{Gerasimov:1965et,Drell:1966jv} sum rule,
    \be
	\lim_{Q^2\to 0}
    \frac{8m_p^2}{Q^2} \int_0^{x{\rm th}} dx \, g_1(x,Q^2)
        = -\kappa^2_p   \,;
    \ee

\noindent $\kappa_p = F_2(0)$ is the proton anomalous magnetic
moment.

Further regarding $\Delta_2$, there is little $g_2$ data for the
proton---there is some from SLAC E155 at higher
$Q^2$~\cite{Anthony:2002hy} and there is some Jefferson Lab Hall C RSS
data at $Q^2 = 1.3$ GeV$^2$ under analysis~\cite{RSS}. Hence, the $g_2$
results rely on models. However, the $g_2$ contributions to the
polarizability corrections are small because the weighting factor
$\beta_2$ in the $\Delta_2$ integral is generally small within the
integral. The weighting factor $\beta$ by contrast is on average close
to 1.







Our main results for $\Delta_{pol}$ are detailed in Table~\ref{table1}.
We evaluated $\Delta_{\rm pol}$ using two different fits to
$g_1$ and $g_2$ and several different parameterizations of
$F_2(Q^2)$.

The current best and most up-to-date parameterization is the one
developed by CLAS EG1~\cite{Fatemi:2003yh, models}. This fit begins with the
most recent published data~\cite{Fatemi:2003yh}, which has $Q^2 \gtrsim
0.15$ GeV$^2$, and is based on AO~\cite{burkert} and MAID~\cite{Drechsel:1998hk}
parameterizations of the resonances, the E155
fit~\cite{Anthony:2000fn} in the deep-inelastic scattering (DIS)
region, and the Wandzura-Wilczek~\cite{Wandzura:1977qf} form $g_2^{WW}=
-g_1 + \int_x^1 g(y)dy/y$ for $g_2$ in the DIS region.  This fit also
gives a good account of the new data~\cite{deur}, which has $Q^2$ down
to 0.05 GeV$^2$. The other structure function fit is that of Simula
{\it et al.}~\cite{Simula:2001iy}, which is based on data available
through the year 2001.  Our results using this fit are shown in the
last column of Table~\ref{table1}, for one $F_2(Q^2)$.

For $F_2$, we show results using the parameterizations of
Kelly~\cite{Kelly:2004hm} and of Ingo Sick~\cite{Sick:2003gm}, and
include the dipole $F_2(Q^2)$ as  a benchmark.  Although the dipole fit is a traditional standard, it does not fit modern
data well and results obtained from it are not reliable.   The Kelly
parameterization fits form factor data well overall, and for $G_E$
elects to fit the polarization transfer results~\cite{Gayou:2001qd}.
(There has been a discrepancy between the Rosenbluth and polarization
transfer determinations of the proton $F_2$.  Theoretical analysis~\cite{Blunden:2003sp} is resolving this by
suggesting that two-photon corrections to the Rosenbluth determination at non-zero $Q^2$
will give agreement with the polarization method.) The continued
fraction parameterization of~\cite{Sick:2003gm} concentrates on the low
$Q^2$ data, and is valid for $0 \le Q ^2\le 0.62$ GeV$^2$.
Beyond this, the $F_2(Q^2)$ contributions to the integrals are small
and we substituted the dipole form. This procedure was also used by
Friar and Sick~\cite{Friar:2003zg} in their analysis of the Zemach
radius.   Substituting the Kelly parameterization
instead made no difference on a scale set by our uncertainty limits.

We show our numerical results in Table~\ref{table1}.  We have
split the $Q^2$ integration into segments [0--0.05]
GeV$^2$ (where no data exist), [0.05--20] GeV$^2$, and $>20$ GeV$^2$,
and show contributions from these regions separately.  We also
separate, except for the lowest $Q^2$ region for $\Delta_1$,
the contributions from $F_2$, $g_1$, and $g_2$.

Three columns in Table~\ref{table1} use the CLAS EG1
model \cite{Fatemi:2003yh} for $g_1$ and $g_2$.  Errors were assigned to
the $F_2$ contribution to $\Delta_1$ using the parameter uncertainties
that coherently gave the largest error.  For $Q^2<0.05$ GeV$^2$
this error was added in
quadrature with a 10\% overall systematic uncertainty as a guess about
the absolute accuracy of the data at low $Q^2$.  The
contribution from $g_1$ was given a 10\% error for the deep-inelastic
region ($Q^2>20$ GeV$^2$), and a 50\% error ($0.05<Q^2<20$) dominated
by the uncertainties in the preliminary CLAS data near $Q^2=0.05$
GeV$^2$.  The errors on $\Delta_2$ were taken to be a conservative
100\%, since there are no significant systematic measurements of $g_2$ at
low $Q^2$.    To combine errors, we grouped the errors due to $F_2$, to $g_1$, and to $g_2$ into three sets, combined the errors within each set coherently, and then combined the errors from the three sets in quadrature.

When doing the $Q^2 < 0.05$ GeV$^2$ part of the $\Delta_1$ integral with the EG1 parameterization, we ensured the GDH cancellation by doing a 
Taylor expansion in $Q^2$.  In terms of the moments
\begin{equation}
\Gamma_{1,2}^{(N)}(Q^2)  = \int_0^{x_{\rm th}} x^N g_{1,2}(x,Q^2) dx \,.
\end{equation}
(which at low $Q^2$ satisfy $\Gamma^{(N)}_{1,2}\sim
(Q^2)^{N+1}$) one can expand Eq.~(\ref{eq:B}) and obtain 
$B_1 = \Gamma_1^{(0)} - 10m_p^2\Gamma_1^{(2)}/(9Q^2) +...$. From the
generalized forward spin polarizability~\cite{vdhaeghen2},
\begin{equation}
\gamma_0(Q^2) = \frac{16\alpha m_p^2}{Q^6}\int_0^{x_{\rm th}} x^2
\left(g_1 - \frac{4m_p^2 x^2}{Q^2}g_2 \right)dx  \,,
\end{equation}
one obtains $\Gamma_1^{(2)}\to \gamma_0 Q^6/(16\alpha m_p^2)$ as
$Q^2\to 0$, and from experiment~\cite{vdhaeghen1} 
$\gamma_0 = [-1.01\pm 0.08$ (stat) $\pm 0.10$ (syst)]$\times 10^{-4}$ fm$^4$.  For the lowest moment, we let
$\Gamma_1^{(0)} = -\kappa_p^2Q^2/(8m_p^2) + c_1 Q^4 + \ldots$ at low
$Q^2$,  and integrate the low $Q^2$ part of $\Delta_1$ keeping terms to 
${\cal O}(Q^4)$.  This gives
\be
\Delta_1[0,Q_1^2] = \left\{ -\frac{3}{4} r_P^2 \kappa^2_p + 18 m_p^2 c_1
		- \frac{5m_p^2}{4 \alpha} \gamma_0 \right\} Q_1^2  \,,
\ee
where $r_P$ is the Pauli radius of the proton.
The constant in front
of the $Q^2$ term in $\Gamma_1^{(0)}$ reflects the GDH sum rule. 
Calculations in chiral perturbation theory~\cite{Ji:1999pd} give $c_1 =
3.89$ GeV$^{-4}$, whereas a fit of order $Q^6$ to $\Gamma_1^{(0)}$ using
the latest CLAS data~\cite{deur} in the range [0.05--0.30] GeV$^2$
yields $c_1 = 2.95\pm 0.11 ({\rm stat})$
GeV$^{-4}$.   We used the experimental value 2.95 GeV$^{-4}$ for Table~\ref{table1}.

Regarding $\Delta_2$, if we assume $g_2 = g_2^{WW}$ we find that $\Gamma_2^{(N)}
= -N\Gamma_1^{(N)}/(N+1)$.  Therefore, if we naively extrapolate this
relation to low $Q^2$, we get $\Delta_2[0,0.05]=-0.40\pm 0.05$
using $\gamma_0$ from above.  From Ref.~\cite{vdhaeghen2}, keeping terms to 
${\cal O}(Q^4)$, we find that 
$\Delta_2[0,Q^2_1]=3m_p^2Q^2_1(\gamma_0-\delta_{LT})/2\alpha$.   Using
the MAID $\pi$-channel estimate~\cite{vdhaeghen2} $\delta_{LT}=1.35\times 10^{-4}$ fm$^{-4}$,
we obtain $\Delta_2[0,0.05]=-1.4$.
Both of these values would lower the final $\Delta_{\rm pol}$.  We quote the
CLAS EG1 model ($-0.24\pm 0.24$) in Table I.

\begin{table*}

\begin{ruledtabular}

\begin{tabular}{ccccccc}
&&&\multicolumn{3}{c}{CLAS EG1}
&
  \multicolumn{1}{c}   {Using Simula {\it et al.} $g_1,g_2$ fit}  \\
term  &  $Q^2$ (GeV$^2$)  & from & Kelly's $F_2$
                                        & Sick's $F_2$&
    dipole & Kelly's $F_2$ \\
\hline \hline
$\Delta_1$& [0,0.05] & $F_2$ and $g_1$ & $0.45 \pm 0.30$ &$ 0.49\pm 0.30$
& $0.60\pm 0.28$ & $-1.78\pm 0.6$     \\
      &   [0.05,20] &  $F_2$  & $7.01 \pm 0.22$  & $6.86 \pm 0.27$  & 7.12 &
           $ 7.01 \pm 0.22$    \\
      &        &  $g_1$    & $-1.10 \pm 0.55$   & $-1.10 \pm 0.55$  & $-1.10 \pm 0.55$&
            $-1.78\pm 1.86$       \\
 & [20,$\infty$]   &  $F_2$ &  0.00 & 0.00   & 0.00  &
      0.00    \\
      &     &  $g_1$    &  $0.12\pm 0.01$ & $0.12 \pm 0.01$ & $0.12 \pm 0.01$ &
       $0.10\pm 0.01 $       \\
\hline
total $\Delta_1$    &   &  & $6.48 \pm 0.89$ & $6.38\pm 0.92$
    & $6.74\pm 0.84$ &   $3.55\pm 2.48$     \\
\hline \hline
$\Delta_2$  & [0,0.05]  & $g_2$  & $-0.24\pm 0.24$
    & $-0.24\pm 0.24$ & $-0.24\pm 0.24$ &
     $-0.72\pm 0.14$       \\
    & [0.05,20] & $g_2$ & $-0.33\pm 0.33$ & $-0.33\pm 0.33$  & $-0.33\pm 0.33$ &
       $-1.14\pm 0.23$       \\
      & [20,$\infty$] & $g_2$  & $0.00$ & $0.00$ & $0.00$ &
         0.00     \\
\hline total $\Delta_2$ &&& $-0.57\pm 0.57$ & $-0.57\pm 0.57$ & $-0.57\pm 0.57$ 
	&   $-1.86\pm 0.37$        \\
\hline \hline
$\Delta_1+\Delta_2$ &&& $5.91 \pm 1.06$ & $5.81\pm1.08$ & $6.18\pm 1.02$ &  $ 1.69\pm 2.51$ \\
\hline
$\Delta_{\rm pol}$ &&& $1.34 \pm 0.24$ ppm & $1.32\pm 0.24$ ppm & $1.40\pm 0.23$ ppm
     & $0.38 \pm 0.57$ ppm \\
\end{tabular}

\end{ruledtabular}

\caption{Contributions to $\Delta_{\rm pol}$ using various models.}
\label{table1}

\end{table*}

For the low $Q^2$ contributions in the Simula {\it et al.}~\cite{Simula:2001iy} fit, we straightforwardly integrated the analytic form.  The numerical differences from the EG1 result arise because a Taylor expansion of $\Gamma_1^{(0)}(Q^2)$ from Ref.~\cite{Simula:2001iy}
leads to a small curvature parameter $c_1$.




For all $Q^2$, the $g_1$ and $g_2$ fits of~\cite{Simula:2001iy} have available 
error bands for the
$g_i$.   To evaluate uncertainties in the polarizability
corrections due to uncertainties in the $g_i$, we recalculated
the $\Delta_i$ using consistently the largest $g_i$ and
smallest $g_i$.  This gives an uncertainty estimate for
$\Delta_2$ from $g_2$ of $\pm 0.37$ units, as quoted in
Table~\ref{table1}.  The uncertainty estimate for $\Delta_1$
due to $g_1$ and $F_2$ is $\pm 2.48$ units.

The uncertainty limits involving $g_2$ on the proton may
appear remarkable considering the amount of existing data.
However, one expects that much of $g_2$ is due to the
Wilczek-Wandzura term, which is gotten directly from $g_1$.
This can be verified from the much larger body of data for
$g_2$ on the neutron (using polarized $^3$He
targets).




Table~\ref{table2} shows $\Delta_Z$ and the Zemach radius calculated
from the form factor sets used above.   Table~\ref{table2} also gives $\Delta_S({\rm data}) - \Delta_Z({\rm calc})$, {\it i.e.}, the value of $\Delta_{\rm pol}$ ``desired'' for consistency between measurement and proton structure corrections calculated using a given form
factor set.

\begin{table}[ht]

\begin{ruledtabular}

\begin{tabular}{lcccc}
Form factor  & $r_P$ (fm) &  $ r_Z$ (fm) & $\Delta_Z$ (ppm)&
                                            residual (ppm) \\
Kelly~\cite{Kelly:2004hm}  &0.878(15)&  1.069(13)  &  $-41.01(49)$ & 2.43(52)  \\
Sick~\cite{Sick:2003gm}    &0.871(35)&  1.086(12)  &  $-41.67(46)$ & 3.09(49)  \\
dipole                     &0.851    &  1.025      &  $-39.32$ & 0.74  \\
\end{tabular}

\caption{Pauli and Zemach radii and $\Delta_Z$ including $\delta_Z^{\rm rad}$
for several form factors.
The column ``residual'' is $\Delta_S({\rm from\ hfs\ data}) -
\Delta_Z$, {\it i.e.}, the value of $\Delta_{\rm pol}$ that
would make theory and data consistent.  In this paper, we
obtained $\Delta_{\rm pol} = (1.3\pm 0.3)$ ppm using inelastic
polarized electron-proton scattering data.  All uncertainties  are calculated from the fits except for Sick's $r_Z$ and $\Delta_Z$, which come from~\cite{Friar:2003zg}.}

\label{table2}

\end{ruledtabular}

\end{table}


In summary, we quote our best value as
\begin{equation}
\Delta_{\rm pol} = (1.3 \pm 0.3) {\rm\ ppm}\,.
\end{equation}
The earlier value of Faustov and Martynenko~\cite{Faustov:yp}
was $(1.4 \pm 0.6)$ ppm.  It is remarkable that
this value, based on few data agrees with the determination using fits to the
extensive CLAS data set.  We thus corroborate extractions of hadronic quantities from data as done in~\cite{Volotka:2004zu,dupays}.

We should now focus on modern form factor parameterizations
represented by~\cite{Sick:2003gm} or~\cite{Kelly:2004hm}, which fit low $Q^2$ data well. Accepting these and Table~\ref{table2} could lead one to desire a larger
$\Delta_{\rm pol}$ to reconcile theory and experiment for
hydrogen hfs.   The experimentally-determined
$\Delta_{\rm pol}$ differs from the value deduced from the measured hyperfine splittings
by over 4 (6) standard deviations of $\Delta_{\rm pol}$ for Kelly (Sick).

We cannot, in our opinion, anticipate that new proton $g_1$ or $g_2$ data will change the evaluation of $\Delta_{\rm pol}$ by enough to reconcile the proton structure corrections with the measured hydrogen hyperfine splittings.  For example, to get the requisite $\Delta_{\rm pol}$ by changing the curvature parameter we called $c_1$ would require making it about 5 times larger than the value we used.  This is essentially unthinkable given data already available.  A further look at elastic form factor fits that could give a more suitable Zemach radius $r_Z$ could well be warranted.

We thank Sebastian Kuhn, Ingo Sick, Silvano Simula, and Marc Vanderhaeghen for
helpful discussions.
This work was supported by the National Science Foundation
under grants PHY-0245056 (C.E.C. and V.N.) and PHY-0400332 (V.N.), and the Department of Energy under contract DE-FG02-96ER41003 (K.A.G.).

\end{document}